\documentclass[universe,article,accept,pdftex,oneauthor]{Definitions/mdpi}

\firstpage{1}
\pubvolume{1}
\issuenum{1}
\articlenumber{0}
\pubyear{2025}
\copyrightyear{2024}
\externaleditor{Firstname Lastname} 
\datereceived{16 January 2025}
\daterevised{9 February 2025} 
\dateaccepted{14 February 2025}
\datepublished{ }
\hreflink{https://doi.org/} 

\Title{Modern Bayesian Sampling Methods for Cosmological Inference: A Comparative Study}

\TitleCitation{Modern Bayesian Sampling Methods for Cosmological Inference: A Comparative Study}

\Author{{Denitsa Staicova}
}

\AuthorNames{Denitsa Staicova}

       \AuthorCitation{Staicova, D.}

\address[1]{%
{Institute for Nuclear Research and Nuclear Energy,}
 Bulgarian Academy of {Sciences; Sofia 1784, Tsarigradsko shosse 72, Bulgaria}
 dstaicova@inrne.bas.bg}
\abstract{ We present a comprehensive comparison of different Markov chain Monte Carlo (MCMC) sampling methods, evaluating their performance on both standard test problems and cosmological parameter estimation. Our analysis includes traditional Metropolis--Hastings MCMC, Hamiltonian Monte Carlo (HMC), slice sampling, nested sampling as implemented in dynesty, and PolyChord. We examine samplers through multiple metrics including runtime, memory usage, effective sample size, and parameter accuracy, testing their scaling with dimension and response to different probability distributions. While all samplers perform well with simple Gaussian distributions, we find that HMC and nested sampling show advantages for more complex distributions typical of cosmological problems. Traditional MCMC and slice sampling become less efficient in higher dimensions, while nested methods maintain accuracy but at higher computational cost. In cosmological applications using BAO data, we observe similar patterns, with particular challenges arising from parameter degeneracies and poorly constrained parameters.}

\keyword{cosmology; Markov chain Monte Carlo; nested samplers}

\begin{document}

\section{Introduction}

The shift towards Bayesian methods 
has been the driving force behind the progress in cosmology in recent years. As~astrophysical experiments have grown in complexity, the~need to handle multiple parameters with complex correlations and various sources of systematic uncertainty has become crucial~\citep{lewis2002cosmological}. Bayesian inference provides a natural framework for combining different datasets, incorporating prior knowledge, and~marginalizing over nuisance parameters~\citep{trotta2008bayes}, including instrumental parameters that characterize the detectors used in specific observations. Notable examples of such complex parameter spaces arise in cosmic microwave background (CMB) experiments, large-scale structure surveys, and~gamma-ray burst observations, where detector response and systematic effects must be analyzed jointly with the physical parameters of interest. This approach has proven especially valuable in cosmological parameter estimation, where it allows for model comparison and uncertainty quantification in the context of limited, and~non-repeatable, observations of the universe~\citep{planck2020parameters}.

At the heart of modern Bayesian inference lie Markov chain Monte Carlo (MCMC) methods, which have revolutionized our ability to sample from complex posterior distributions. Since their introduction by \citet{metropolis1953equation} and generalization by \citet{hastings1970monte}, MCMC methods have become the foundation for practical Bayesian computation. These methods enable the exploration of high-dimensional parameter spaces and the calculation of marginal distributions that would be impossible or impractical through direct numerical integration. For~a few reviews on the subjects, see~\cite{Buchner:2021kpm, Colgain:2023bge,Albert:2024zsh}.

In this paper, we examine several key samplers used as  crucial tools in cosmological inference. We compare traditional MCMC implementations with more recent developments including Hamiltonian Monte Carlo (HMC), which uses gradient information to improve sampling efficiency; slice sampling, which adaptively determines step sizes; and nested sampling, which simultaneously computes the Bayesian evidence while sampling from the posterior. We also compare them to  another nested sampler provided by the package Polychord. Through a series of test problems and cosmological applications, we evaluate their performance, providing practical guidance for their use in parameter estimation {challenges.}
 {{While these sampling methods have been well studied individually, this work provides a novel comprehensive comparison framework where all methods are evaluated on identical problems using consistent metrics. This unified approach, testing both standard distributions and cosmological likelihoods, offers unique insights into the relative strengths and trade-offs between methods.}
}

The paper is organized as follows. In~Section~\ref{sec2}, we present an overview of the sampling methods we test. Section~\ref{sec3} presents the specific implementations and datasets we use.
Section~\ref{sec4} introduces our suite of test problems, designed to probe different challenging aspects of sampling.
In Section~\ref{sec5}, we apply these samplers to cosmological parameter estimation, focusing on the $\Lambda$CDM model with varying numbers of free parameters. Section~\ref{sec6} concludes our study with a brief overview of new promising~methods.

\section{Overview of Sampling~Methods}\label{sec2}

In this paper we will review several key algorithms used in cosmology. We start by presenting a short overview of the~methods.

\subsection{Traditional~MCMC}
The Metropolis--Hastings algorithm \cite{metropolis1953equation, hastings1970monte} is a foundational approach in Bayesian inference. It generates samples from a target distribution $\pi(\theta)$ using a proposal distribution $q(\theta' | \theta)$. At~each step, a~new state $\theta'$ is accepted with {probability:}

\begin{equation}
    \alpha(\theta' | \theta) = \min\left(1, \frac{\pi(\theta')q(\theta|\theta')}{\pi(\theta)q(\theta'|\theta)}\right).
\end{equation}

For symmetric proposals ($q(\theta'|\theta) = q(\theta|\theta')$), this reduces to the Metropolis ratio. The~algorithm's efficiency depends strongly on the choice of proposal distribution, with~poor choices leading to either high rejection rates or slow exploration of the \mbox{parameter~space.}

\subsection{Hamiltonian Monte~Carlo}
This method, introduced by~\cite{duane1987hybrid}, uses gradient information to improve sampling efficiency. HMC extends MCMC by introducing auxiliary momentum variables $p$ and using Hamiltonian dynamics to propose new states. The~system evolves according to:
\begin{equation}
    H(\theta, p) = -\log \pi(\theta) + \frac{1}{2}p^TM^{-1}p
\end{equation}
where $M$ is a mass matrix. The~dynamics follow Hamilton's equations:
\begin{align}
    \frac{d\theta}{dt} &= M^{-1}p \\
    \frac{dp}{dt} &= \nabla_\theta \log \pi(\theta).
\end{align}

These equations are typically solved using the leapfrog integrator, which preserves volume in phase space. By~incorporating Hamiltonian dynamics, HMC can achieve better exploration of the parameter space, particularly in high~dimensions.

\subsection{Slice~Sampling}
Developed by~\cite{neal2003slice}, slice sampling adaptively determines step sizes, potentially offering better mixing than traditional methods. Slice sampling introduces an auxiliary variable $u$ to sample from an augmented space:
\begin{equation}
    p(\theta, u) \propto \begin{cases}
    1 & \text{if } 0 \leq u \leq \pi(\theta) \\
    0 & \text{otherwise}.
    \end{cases}
\end{equation}

The algorithm alternates between drawing $u \sim \text{Uniform}(0, \pi(\theta))$ and sampling $\theta$ uniformly from the ``slice'' $\{\theta: u \leq \pi(\theta)\}$.
In this way, the~method adaptively adjusts the step~size.

\subsection{Nested~Sampling}
Introduced by~\cite{skilling2006nested}, nested sampling simultaneously computes the evidence and produces posterior samples.  Nested sampling transforms the problem into a one-dimensional integration over prior mass $X$:
\vspace{-6pt}
\begin{equation}
    Z = \int \mathcal{L}(\theta)d\theta = \int_0^1 \mathcal{L}(X)dX,
\end{equation}
where $\mathcal{L}(X)$ is the inverse of the prior cumulative distribution. The~algorithm maintains a set of ``live points'' drawn from the prior, and through increasing likelihood constraints, iteratively replaces the lowest-likelihood point (which becomes a ``dead point''). This provides both posterior samples and the evidence $Z$.

{{The evidence $Z$ plays a crucial role in model comparison, allowing us to compute Bayes factors between competing cosmological models. While traditional MCMC methods require post-processing approximations to compute $Z$, nested sampling provides it directly along with the posterior samples. For~technical details on evidence computation, see Ap}pendix~\ref{sec:evidence}.}

\subsection{PolyChord}
PolyChord is a specialized version of the nested sampling approach developped by~\citep{Handley:2015fda, Handley:2015vkr}. It was designed specifically for the high-dimensional parameter spaces typical in cosmology. It employs slice sampling to generate new points within nested sampling, combining the advantages of both methods. This approach proves particularly effective for parameter spaces with complex geometries and~degeneracies.

These methods differ fundamentally in their approach to sampling: while traditional MCMC and HMC explore the parameter space through chains that converge to the posterior distribution, nested sampling works ``from the outside in,'' systematically moving through nested likelihood contours. Slice sampling stands out for its adaptive nature, requiring minimal tuning while maintaining good~efficiency.

\section{Numerical~Methods}\label{sec3}
\textbf{{Sampling Packages}:}
Our analysis employs several widely used sampling packages,  chosen for their reliability and proper~documentation:
\begin{itemize}
      \item Traditional MCMC: We use PyMC~\citep{pymc} ({\url{https://github.com/pymc-devs/pymc}}, {accessed on 16.01.2025),}
a~ probabilistic programming framework implementing various MCMC algorithms and providing automated initialization~procedures.

    \item HMC: We implement HMC using NumPyro~\citep{Phan:2019elc} ({\url{https://github.com/pyro-ppl/numpyro}}{, accessed on 16.01.2025}), which provides efficient, JAX-based implementations of HMC and No-U-Turn Sampler (NUTS)~\cite{Hoffman:2011ukg}. This choice offers automatic differentiation capabilities that are crucial for HMC while maintaining computational efficiency through just-in-time~compilation.

    \item \textls[-20]{Slice Sampling: Our implementation follows Neal's algorithm~\citep{neal2003slice}, with~adaptations for high-dimensional parameter spaces. The~code incorporates automatic step-size adjustment and implements the stepping-out procedure for slice width determination. {{While slice sampling is implemented in some statistical packages like PyMC, we developed a specialized version optimized for our benchmark framework to ensure consistent interface with other samplers and direct control over the stepping-out~procedure.}}}

   \item \textls[-18]{{emcee} ({\url{https://github.com/dfm/emcee}}{, accessed on 16.01.2025}): {An affine-invariant ensemble sampler, known for its robustness and wide adoption in the astronomy~community}~\citep{Foreman-Mackey:2012any}.}

     \item \textls[-15]{Nested Sampling: We use dynesty~\citep{Speagle:2019ivv, sergey_koposov_2024} ({\url{https://github.com/joshspeagle/dynesty}}{, accessed on 16.01.2025}), a~dynamic nested sampling package designed for astronomical~applications. }

      \item PolyChord~\citep{Handley:2015fda} ({\url{https://github.com/PolyChord/PolyChordLite}}{, accessed on 16.01.2025}): A specialized nested sampling algorithm particularly suited for high-dimensional parameter spaces and for multi-modal~distributions.
\end{itemize}

{{Another popular nested sampler is} MultiNest~\citep{Feroz:2008xx}, which is widely used  in galaxy spectral energy distribution fitting. Here, however, we focus on dynesty and PolyChord as they are better suited for our cosmological likelihood functions.
}

\textbf{{Benchmark framework and data processing:}}
The benchmark framework we implemented  ({the code will be made public upon publication on \url{https://github.com/dstaicova/samplers_benchmark}}, {accessed on 16.01.2025}) standardizes the metrics we track across all samplers while accounting for their differences. To~evaluate the samplers, we  use several performance metrics: runtime, memory usage, effective sample size
(ESS) per second, and parameter~accuracy.

\textbf{{Runtime and memory usage}} directly measures computational cost and memory allocation.  Memory profiling tracks both resident set size (RSS) and virtual memory size (VMS), as~well as possibly considering whether some samplers are parallelized. \textbf{{Effective sample size (ESS) per second}} measures sampling efficiency by calculating how many effectively independent samples are generated per unit time, accounting for autocorrelation in the chains. \textbf{{Init sensitivity}} measures the sensitivity to random initialization fluctuation (i.e., starting the code with different seeds). \textbf{{Parameter accuracy}} assessments utilize deviation from known true values in test problems and consistency checks in cosmological~applications.

The output of each sampler is post-processed to ensure as fair as possible a comparison. For~MCMC methods, we implement burn-in removal, while for~nested samplers, we utilized the reweighted samples. We export the direct samples for each model, without~using additional tools such as getdist. We track the convergence and the  R-hat statistics where applicable and monitor acceptance rates and effective sample sizes. The~framework includes extensive error handling and diagnostic~reporting.

{{Several samplers support parallelization. Nested sampling methods naturally parallelize their likelihood evaluations, with~both dynesty and PolyChord offering MPI implementations, while emcee enables parallel chain execution through its built-in pool feature, and PyMC supports parallel sampling via multiple chains. However, parallelization efficiency can vary significantly depending on likelihood complexity and communication overhead.}}

{{Each method handles burn-in differently: traditional MCMC and emcee use explicit burn-in periods (50\% and 200 steps, respectively, except~for emcee, for which we use 100 steps), HMC incorporates warmup steps for adapting the step size and mass matrix, and nested sampling methods continuously replace points and thus do not require explicit burn-in.}}

\section{Test~Problems}\label{sec4}
\begin{figure}[H]
\includegraphics[width=1\textwidth]{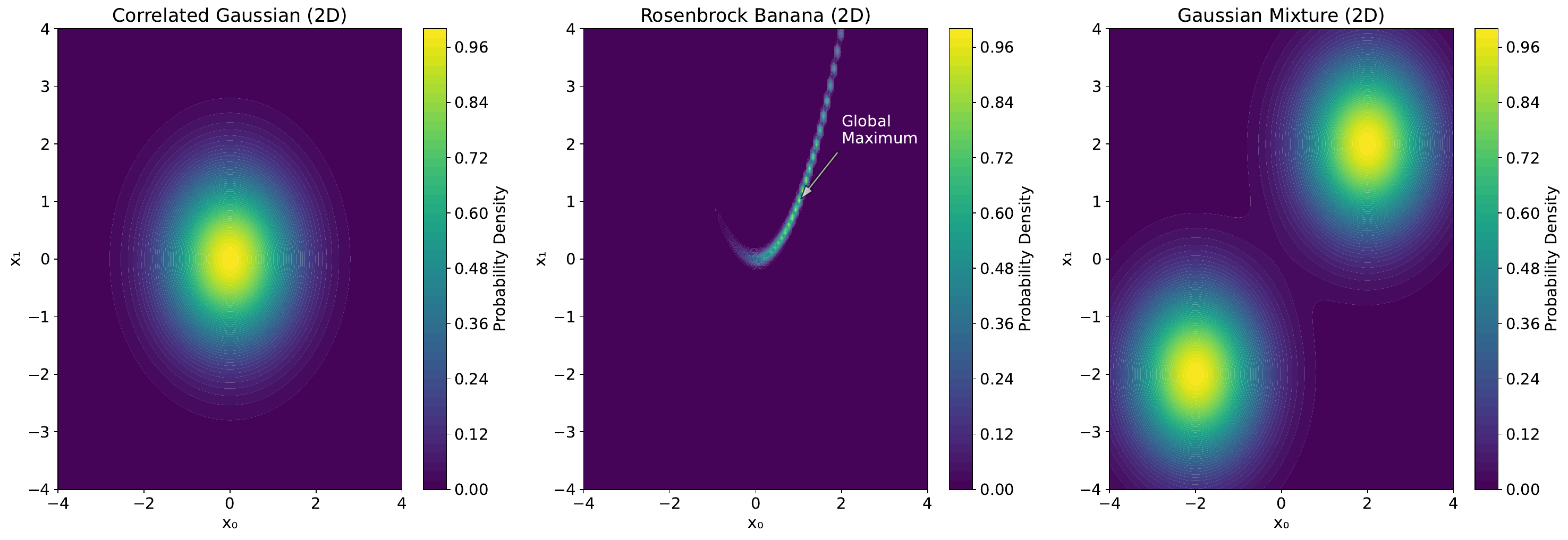}
 \caption{The surface plots corresponding to the three test problems. The~global maximum that the sampler needs to find is marked in the case of the Rosenbrock distribution; the~two others correspond to single and double Gaussians,~respectively.}
    \label{fig:test_problems}
\end{figure}

We benchmark the samplers on a set of test problems, designed to examine different challenges commonly encountered in parameter estimation. The~surface plots of the test problems' distributions can be seen in Figure~\ref{fig:test_problems}.

\textls[-20]{\textbf{{Correlated Gaussian:}}
The simplest test case involves a multivariate Gaussian distribution:}
\vspace{-6pt}\begin{equation}
    \log \mathcal{L}(\mathbf{x}) = -\frac{1}{2}\sum_{i=1}^d x_i^2,
\end{equation}
where $d$ is the dimension of the parameter space and $x_i$ represents the $i$-th parameter~value.

This distribution tests basic convergence properties and scaling with dimensionality. It establishes a baseline  efficiency under ideal conditions, particularly relevant for cosmological applications where approximate Gaussianity often holds near the maximum~likelihood.

\textbf{{Rosenbrock (Banana) Distribution:}}
The Rosenbrock function, also known as the banana distribution due to its characteristic shape, presents a challenging curved degeneracy:
\begin{equation}
    \log \mathcal{L}(\mathbf{x}) = -\sum_{i=1}^{d-1} \left[100(x_{i+1} - x_i^2)^2 + (1-x_i)^2\right].
\end{equation}

This distribution tests samplers' ability to navigate narrow, curved valleys in parameter space, a~feature often encountered in cosmological parameter estimation where parameters exhibit strong non-linear correlations. A~cosmological example of such curved distribution might come from the degeneracy between the mass density and the Hubble constant~\cite{Abdalla:2022yfr}.

\textls[-15]{\textbf{{Gaussian Mixture:}}
A bimodal distribution testing mode-finding and mixing capabilities:}
\begin{equation}
    \log \mathcal{L}(\mathbf{x}) = \log\left[\exp\left(-\frac{1}{2}\sum_{i=1}^d (x_i+2)^2\right) + \exp\left(-\frac{1}{2}\sum_{i=1}^d (x_i-2)^2\right)\right].
\end{equation}

This distribution challenges samplers to properly sample from multiple distinct modes, a~situation that arises in various cosmological contexts, like gravitational lensing, neutrino mass hierarchy, and modified gravity theories (for example~\cite{He:2024udi, ajani2020constraining, Lopez:2021ikt}).

\subsection{Performance Metrics and~Accuracy}

To evaluate the accuracy, we use two complementary accuracy metrics: mean-based and distribution-based, evaluating both the ability to find the correct parameter values and to properly explore the target distribution (see the Appendix for details). All metrics are normalized within each test problem to facilitate comparison across different distributions and~dimensions.

\subsection{Results}
\begin{figure}[H]
\includegraphics[width=1\textwidth]{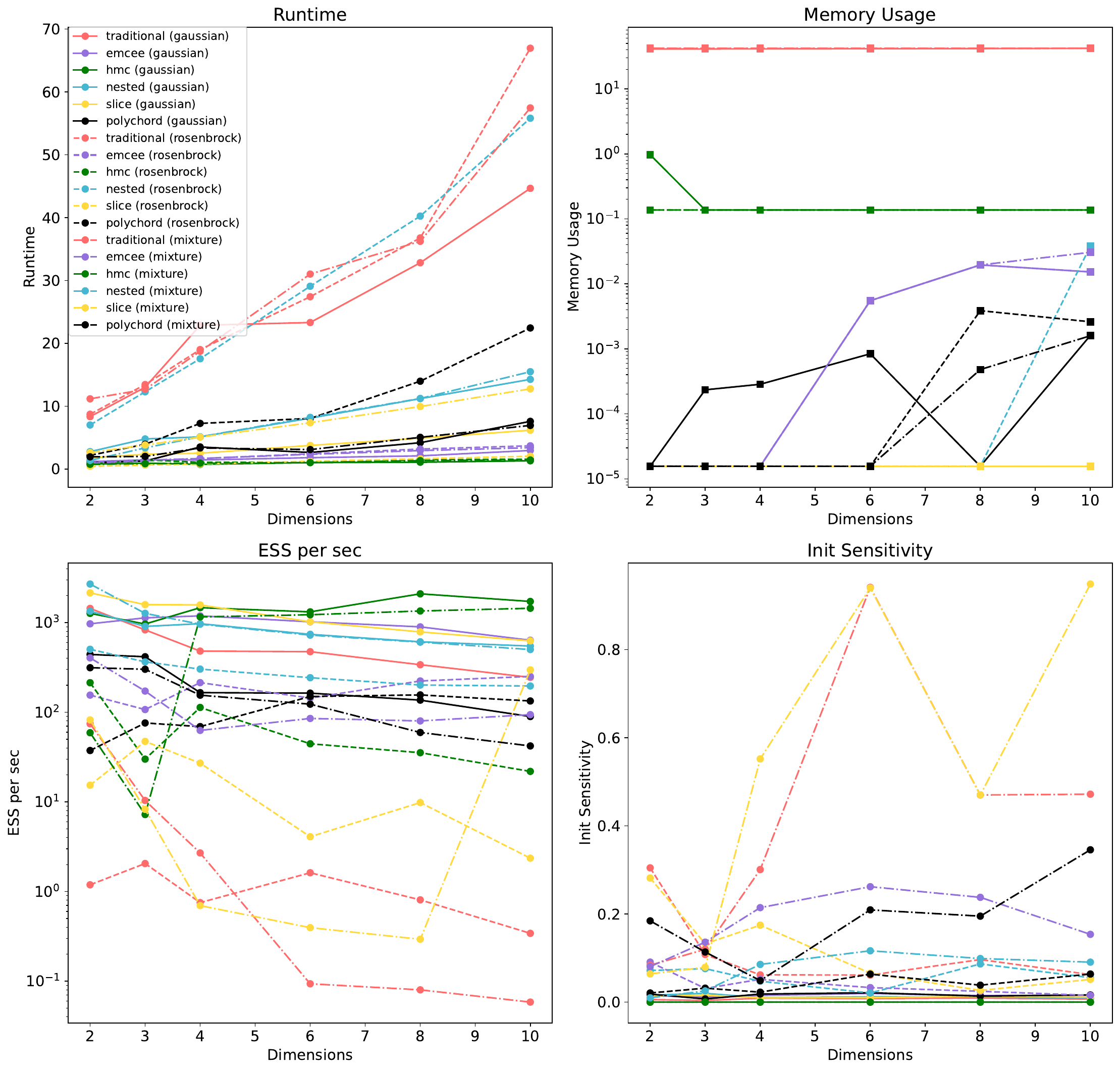}
 \caption{{The summary of the metrics}
 we track for the different samplers. We show here the runtime, the~memory usage, the~ESS per sec, and the Init~sensitivity. }
    \label{fig:test_problems_runs}
\end{figure}

\begin{figure}[H]
\includegraphics[width=1\textwidth]{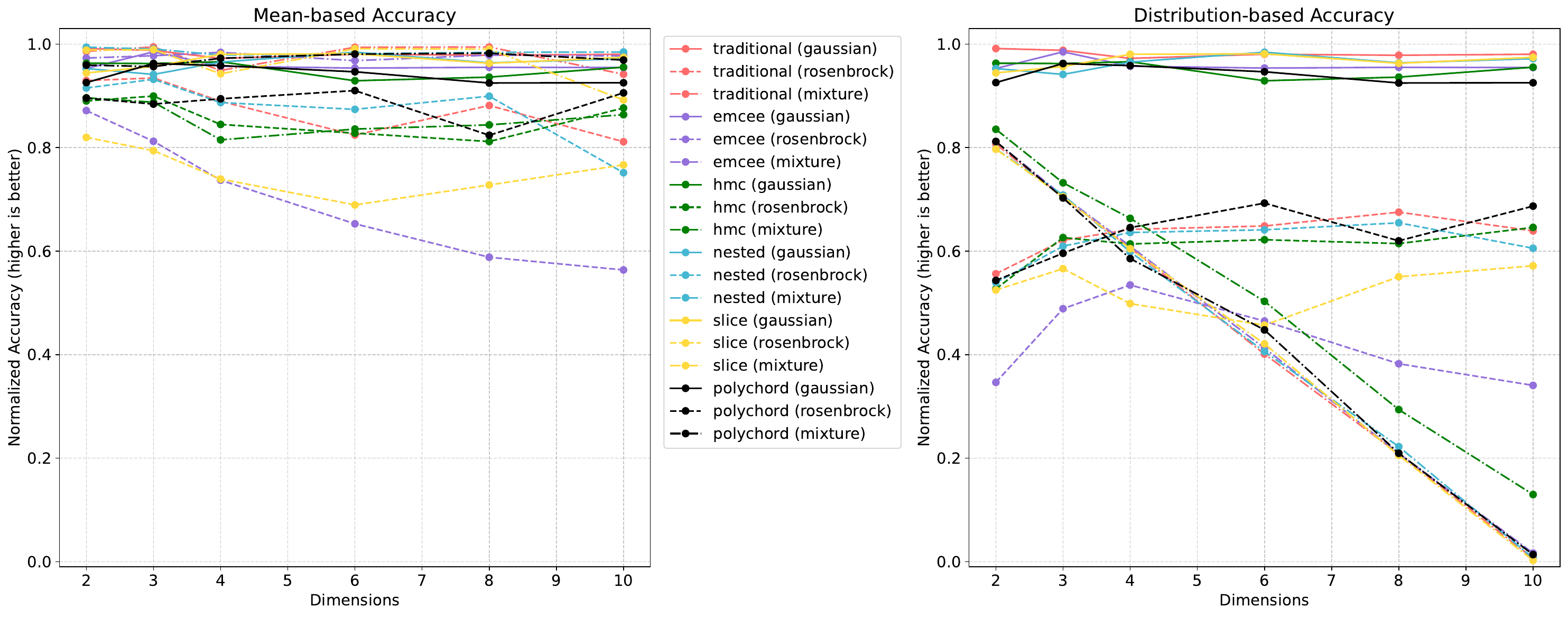}
 \caption{Comparison between the mean accuracy and the distribution accuracy for different samplers. The~normalization is described in the~{Appendix A.}
}
    \label{fig:test_problems_accuracy}
\end{figure}

The results are shown in Figures~\ref{fig:test_problems_runs} and~\ref{fig:test_problems_accuracy}. All samplers perform well on the simple Gaussian case, with~HMC and slice showing particularly strong performance. The~mean-based and distribution-based metrics align closely, as~expected for a unimodal, symmetric distribution. Runtime scaling with dimension  for most samplers is modest, with~slice sampling, HMC, nested, and emcee showing practically the same efficiency in terms of ESS per~second.

We focus our analysis on the described above  key metrics: runtime and memory usage, effective sample size (ESS) per second, and Init~sensitivity.

\unskip

The Rosenbrock function reveals significant differences between samplers. Its curved ”banana” shape poses challenges for traditional MCMC and slice methods, evidenced by their declining ESS per second with increasing dimensions. While most samplers maintain reasonable mean accuracy, the~distribution-based metric shows larger discrepancies, indicating difficulty in properly exploring the curved parameter space. Here, slice and emcee fare worst in terms of accuracy, while the ESS per sec show poor performance for traditional MCMC, slice, and~HMC.

The bimodal distribution presents the most challenging test case, with~substantial variations in sampler performance. The~mean-based accuracy becomes less meaningful here, as~the true mean lies between modes. The~distribution-based accuracy reveals significant degradation with the increase of dimensions, due to increased probability of the sampler jumping between modes. Here, HMC shows a bit better performance in terms of accuracy and ESS per sec, with a very slight cost in terms of runtime but requiring more memory. {{We also tested PolyChord with its clustering option enabled, designed for multimodal distributions, but~found only minimal improvement in the accuracy and a slight increase in the runtime. Since it sets the accuracy between HMC and PolyChord without clustering, it seems that the base algorithm already handles multiple modes adequately.}}

It is important to note that some metrics require careful interpretation: the ESS measurements, designed for MCMC methods, do not fully capture nested sampling efficiency, better characterized by the ratio of live to dead points at each likelihood threshold. For~example, for~simple Gaussian distributions, PolyChord maintains reliable accuracy but with lower ESS per second compared to traditional MCMC methods. In~the Rosenbrock case, its performance remains stable across dimensions, effectively navigating the curved parameter space due to its slice sampling component. On~the other hand, the~traditional method was convergent only for the Gaussian distribution. We tried increasing the sample size and the tuning, but~that only led to an increase of runtime and memory, not better~convergence.

The Init sensitivity metric is the final metric we track. It tracks the samplers' reliability across different random seeds---a key consideration for reproducible cosmological analyses. While most samplers show minimal sensitivity at low dimensions (<0.005 for 1D and 2D), we see increasing variability in higher dimensions. This effect becomes more pronounced with the decrease of the accuracy, for~example for the mixture problem. HMC demonstrates notably stable performance across dimensions, likely due to its geometric properties, while traditional MCMC shows a moderate increase in sensitivity with~dimensionality.

{{In summary, each sampler shows distinct performance characteristics across our test problems. Traditional MCMC and slice sampling excel in low dimensions but struggle with efficiency in higher-dimensional spaces. HMC maintains consistent performance across dimensions, particularly evident in the Rosenbrock case, though~at increased computational cost. Nested samplers (both dynesty and PolyChord) demonstrate superior reliability for multimodal problems, while emcee shows strong ESS per second performance but with increased parameter uncertainties in higher dimensions. These patterns suggest that the choice of a sampler should be based on the specific features of the problem distribution, such as dimensionality,  multimodality, or curved degeneracies.}}

\section{Cosmological~Applications}\label{sec5}
\subsection{Marginalized BAO~Likelihood}

The code uses the marginalized likelihood that uses data from Baryonic acoustic oscillations (BAO), in~this case, the~newest DESI results~\cite{DESI:2024mwx}, for~which the dependence on the Hubble constant ($H_0$) and the sound horizon ($r_d$) are integrated out. The~details on this method can be found in~\cite{Staicova:2021ntm, Benisty:2024lmj}, and we will leave them out for brevity. The~marginalized BAO likelihood provides a good test case due to the fact it has been tested already on more extensive numerical datasets, and it also exhibits some of the features of cosmological inference problems---for example, the marginalization process might introduce non-Gaussianity, while the dark energy parameters might have curved degeneracies~\citep{taylor2010analytic, gong2014effect}.

\begin{figure}[H]
\includegraphics[width=0.85\textwidth]{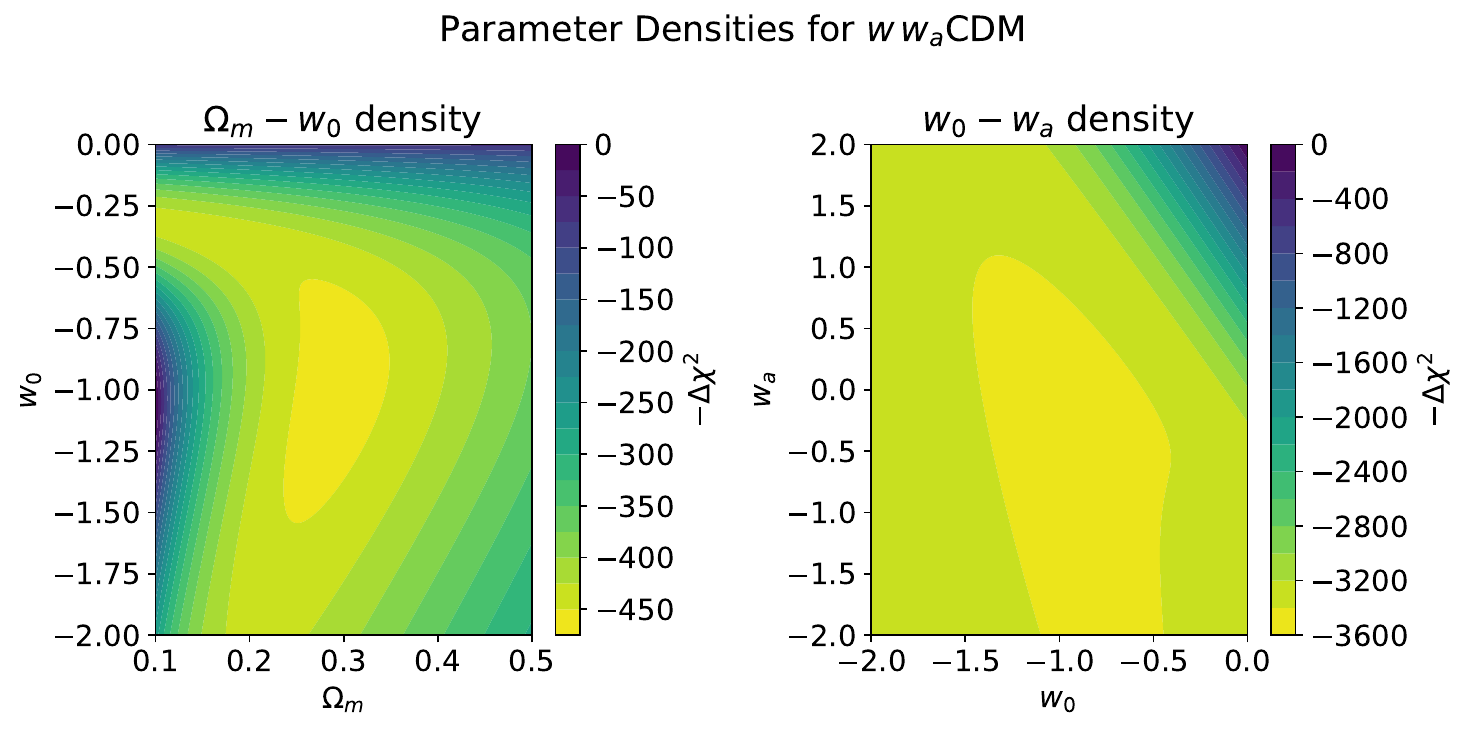}
 \caption{A slice of the density of the likelihood of the $ww_a$CDM model when two parameters vary and the other is set to its fiducial value $\Omega_m=0.3, w_0=-1, w_a=0$. Note that this is not a contour plot of the posterior but a direct evaluation of the likelihood~function.}
    \label{fig:BAO_likelihood}
\end{figure}

{{The final likelihood depends only on the matter density} $\Omega_m$, {the~curvature density} $\Omega_K$, {and the dark energy parameters} $w(a)=w_0 + w_a(1-a)$ corresponding to the CPL parametrization~\citep{Chevallier:2000qy, Linder:2005ne} (where $a$ is the scale factor).}
 We implemented the six samplers discussed above on the cosmological likelihood and tested each of them on the BAO likelihood with increasing dimensionality (1--4 dimensions), corresponding to different cosmological models: $\Lambda$CDM (1D) with $\Omega_m$ as a free parameter; $\Omega_K$CDM, which adds spatial curvature  $\Omega_K$ (2D); dark energy model $ww_a$CDM (3D), for which the free parameters are $\Omega_m, w_0, w_a$; and $\Omega_Kww_a$CDM (4D), which combines all the four parameters. In~the study below, we use the following priors: $\Omega_m\in (0.1,0.5), \Omega_K\in(-0.3,0.3),w \in (-2,0), w_a\in(-2,2)$, applied to the respective models{.}
{{The structure of the BAO likelihood reveals why our test distributions are relevant for cosmological sampling. Direct visualization of the likelihood surfac}e (Figure \ref{fig:BAO_likelihood}) shows curved parameter degeneracies, particularly in the $\Omega_m$-$w_0$ and $w_0$-$w_a$ planes, motivating our use of the Rosenbrock test distribution. While this BAO likelihood exhibits a single peak with curved degeneracies, cosmological applications can encounter genuinely multimodal distributions that the Gaussian mixture models simulate---for example in modified gravity theories with distinct viable solutions.}

\subsection{Performance~Analysis}

\begin{figure}[H]
\includegraphics[width=1\textwidth]{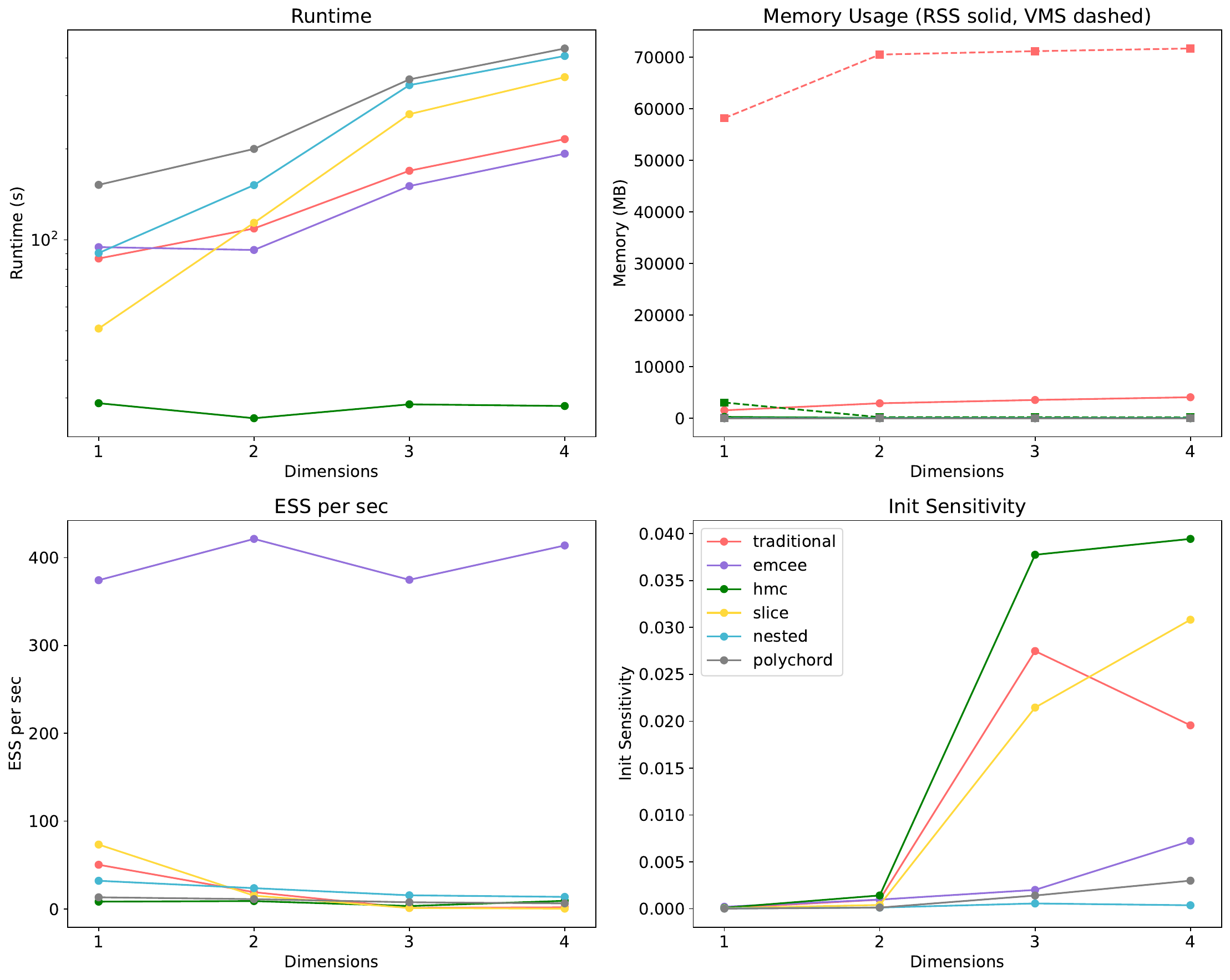}
 \caption{The summary of the benchmark on cosmological models using the different~samplers.}
    \label{fig:BAO_metrics}
\end{figure}

The performance metrics can be seen on Figure~\ref{fig:BAO_metrics}. Runtime scaling follows approximate power laws with dimension, but~coefficients vary substantially between methods. For~example, in~the $ww_a$CDM case (3D), typical runtimes range from $\sim$30 s for HMC to $\sim$340 s for PolyChord (with nested doing a little bit better but close to it), with~traditional MCMC and emcee falling in between at $\sim$150 s, and slice at about $\sim$250 s. The~scaling with dimension is approximately constant for HMC (due to its gradient-based updates), while nested sampling methods show steeper, approximately quadratic scaling due to the increasing difficulty of sampling from the constrained prior volume. Still, they maintain reliable exploration of the parameter space and provide the calculated evidence, which facilitates further analysis. Slice demonstrates very steep scaling with dimensions. In~terms of runtime, HMC does best, while PolyChord does~worst.

Memory requirements scale distinctly for each sampler implementation. Traditional MCMC shows significant memory use, while HMC maintains relatively stable memory usage through its computational graph optimization, but~at the cost of requiring JAX-compatible likelihood functions. Nested sampling methods present unique memory profiling challenges due to their parallel implementation structure, potentially leading to underestimation in our measurements as the memory allocation occurs across multiple processes. {{The observed memory usage drops with increasing dimensionality for some samplers reflect their adaptive memory management strategies rather than numerical instabilities. For~example, nested samplers may adjust their internal storage based on active parameter count or employ a dedicated garbage collector, while MCMC methods can benefit from more efficient memory allocation in sparse parameter spaces}.}

Sampling efficiency, measured through ESS per second (Figure \ref{fig:BAO_metrics}), demonstrates strong dependence on both dimensionality and parameter type. HMC maintains consistent efficiency up to 3D before showing some decline in 4D, likely due to the increasing complexity of the parameter space affecting its momentum updates. Traditional MCMC shows stable but lower efficiency, while nested samplers achieve consistent performance despite lower raw sampling rates. Slice sampling shows a notable deterioration in performance with~dimensionality.

In this case, the Init sensitivity demonstrates more significant scaling with dimensions, likely related again to the increase of errors in the constraints. We see that it is quite low in 1D, but~for $d>2$, it quickly increases. Here, the nested samplers and emcee show the most stable~performance.

{{Overall, our cosmological application reveals trade-offs between computational efficiency and sampling reliability. HMC provides the best balance of runtime and accuracy, particularly for higher-dimensional models,} though~gradient information is required. Traditional MCMC and emcee offer reasonable performance for simpler models ($\Lambda$CDM, $\Omega_K$CDM) but show diminishing efficiency with additional parameters. Nested sampling methods, while computationally intensive, prove valuable for exploring complex parameter spaces and providing evidence calculations. The~increased initialization sensitivity in higher dimensions suggests that multiple runs may be necessary for robust results, particularly when exploring extended cosmological models.
}

\subsection{Accuracy~Analysis}

\begin{figure}[H]
\includegraphics[width=1\textwidth]{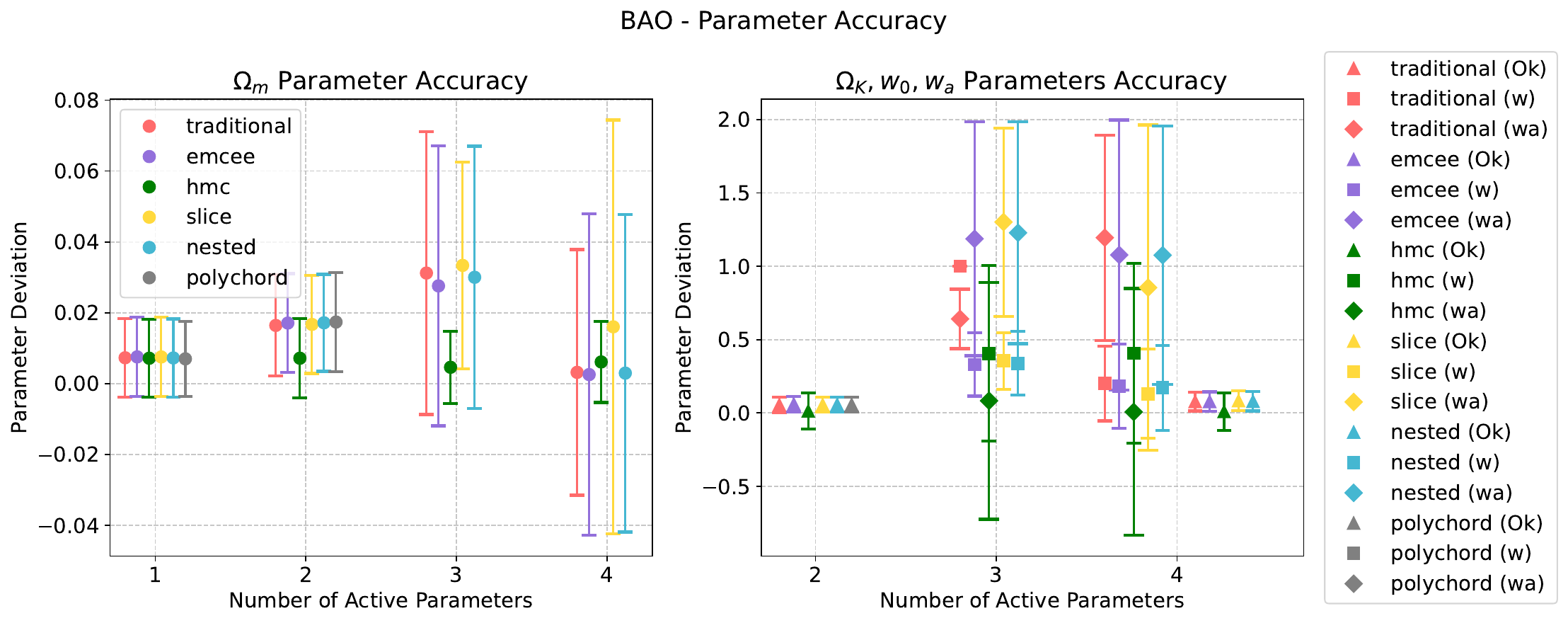}
 \caption{The left panel shows the well-constrained $\Omega_m$ for all models. The~right panel displays deviations for the additional parameters: spatial curvature ($\Omega_k$) in the 2D model and the 4D model, equation of state $w$ in the 3D model, and~both $w$ and $w_a$ in the 4D model.
}
    \label{fig:BAO_accuracy}
\end{figure}

Parameter constraint quality, shown in Figure~\ref{fig:BAO_accuracy}, varies significantly between cosmological parameters. $\Omega_m$ shows consistent accuracy across all samplers and dimensions,
with deviations from the fiducial value ($\Omega_m = 0.3$) typically around $0.007\pm 0.002$. This indicates robust sampling of well-constrained parameters regardless of method choice. We also note that the constraint degrades with the increase of dimensions, and also that HMC is particularly good in higher dimensions in terms of~accuracy.

The spatial curvature has a small deviation from the fiducial value but with large standard deviation $\Omega_K=0$ of $0.05 \pm 0.05$, with only HMC giving a negative mean with higher error ($\Omega_K=-0.013\pm 0.12$).

Dark energy parameters prove more challenging, showing significantly larger uncertainties and stronger parameter degeneracies, with $w$ constrained much better than $w_a$. {{Using the fiducial values of }$w_0=-1$ and $w_a=0$,} for the $3D$ case, we see $w_0=-0.67\pm 0.3$, while the range on $w_a$ is much bigger, $w_a\approx -1 \pm 0.8$. Finally, in~our test of $\Omega_Kw_0w_a$CDM, we again see large errors, especially in $w_a$, but surprisingly tight constraints on $\Omega_K$.

The {{analysis of the relationship between}} $w_0$ and $w_a$ particularly highlights differences between samplers, with~nested methods
showing advantages in exploring the degenerate parameter space. This pattern reflects the inherent degeneracies in these parameters, the~increased difficulty in constraining them as the parameter space expands, and the decreased sensitivity of the marginalized likelihood to~them.

In conclusion, all samplers maintain comparable performance levels, with~no method showing significant advantages in accuracy, though~traditional MCMC and emcee exhibit slightly larger uncertainties in the higher-dimensional~cases.

\subsection{Implementation~Challenges}

Implementing the benchmark for cosmological likelihood showed a few significant challenges. The~memory was hard to measure due to the rapid allocation and deallocation of memory and the parallel processing of nested samplers. Furthermore, creating a uniform interface across samplers was a major problem, since it required careful handling of different parameter space representations (unit cube vs. physical space), various input and output formats (for example normal likelihood vs. JAX-compatible) and chain structures, and divergent initialization procedures. Consequently, the~code required extensive error handling  due to numerical instabilities in likelihood evaluations and different convergence behaviors. Finally creating the diagnostics was not trivial. This is  because not all samplers provide the same diagnostic metrics (for example, we could not obtain the acceptance rate for HMC), some diagnostics (like R-hat) require multiple chains, and ESS calculations differ between MCMC and nested sampling approaches, as seen in Figures~\ref{fig:test_problems_runs} and~\ref{fig:BAO_metrics}.

{{The impact of error handling varies across samplers. Our implementation includes basic error checking for} parameter consistency and numerical stability, though~the internal error handling and diagnostics of each sampling package may have additional computational costs that are not directly measured in our benchmarks. These diagnostic routines are integral to ensuring reliable results, but their precise computational impact would require more detailed profiling.}

The results demonstrate that while all methods achieve similar accuracy for well-constrained parameters, their performance diverges significantly for parameters with strong degeneracies. This suggests that method selection
for cosmological applications should consider both the specific parameters of interest and computational resource~constraints.

\section{Discussion and~Outlook}\label{sec6}

Bayesian inference in cosmology often requires sampling from complex probability distributions in high-dimensional parameter spaces. {{Through our unified benchmarking framework, comparing multiple samplers simultaneously on identical problems, we find that the choice of sampling method can significantly impact both the accuracy of results and computational efficiency.}}
We have examined several of the most basic MCMC methods and tried to benchmark them both in very simple test problems and in realistic cosmological settings. We designed the code using well-known Python libraries, facing numerous challenges related to the simultaneous work of all the samplers together. We show that no single method is significantly better than the others, and all of them can be used in certain~situations.

In particular, traditional MCMC maintains reliable performance but shows limitations in higher dimensions or in complicated distributions. The~strong performance of slice
sampling in low dimensions suggests simpler methods can still be useful, though~one needs to take into account the deterioration of their performance  in higher~dimensions.

The success of emcee in terms of ESS per second as it combines with increased parameter uncertainties in higher dimensions suggests that raw sampling efficiency does not necessarily indicate optimal posterior exploration. This
observation particularly impacts cosmological applications, where accurate
uncertainty estimation proves crucial. The~nested samplers---nested (dynesty) and PolyChord---on~the other hand, while keeping modest ESS per sec, demonstrated good accuracy in the accuracy~metrics.

To conclude, we would like to discuss some alternatives to the simple samplers considered here. Recent years have seen significant advances in sampling methods, particularly those addressing the unique challenges of cosmological applications. These developments broadly fall into several categories, each offering novel approaches to overcome specific limitations of traditional~methods.

Neural-based approaches have emerged as a promising direction, with~methods like the neural sampling machine~\citep{dutta2022neural} utilizing synaptic noise for learning and inference to approximate Bayesian inference. This neural acceleration is particularly relevant for cosmology, where complicated instrumental likelihoods can be computationally expensive. In~cases where likelihoods are intractable, likelihood-free inference methods employing neural density estimators~\citep{alsing2019fast, jeffrey2021likelihood} have proven to be powerful alternatives to traditional MCMC~approaches.

For multi-modal distributions, common in cosmological applications, several innovative methods have been developed. Parallel tempering~\citep{sambridge2014parallel} has shown improved efficiency in exploring such distributions, while continuous tempering~\citep{graham2017continuous} introduces a continuous temperature parameter for smoother transitions between tempered distributions. The~challenge of quasi-ergodicity in Monte Carlo simulations has been addressed through various approaches~\citep{neirotti2000approach, frantz1990reducing}, including the use of optimal transport theory for designing more efficient proposal~distributions.

Geometric considerations have also driven significant methodological advances. Orbital MCMC~\citep{neklyudov2022orbital} uses principles from Hamiltonian mechanics to preserve geometric structures in the target distribution, while geometric HMC variants~\citep{betancourt2015hamiltonian, betancourt2018geometric} offer improved adaptation to the target distribution's geometry through higher-order differential geometric structures. These methods show particular promise for high-dimensional problems with complex geometries, though~their practical implementation often requires significant computational resources and~expertise.

The integration of machine learning with sampling methods has opened new avenues for improvement. Adaptive Monte Carlo augmented with normalizing flows~\citep{gabrie2022adaptive} combines normalizing flows with MCMC methods, using learned transformations to enhance sampling efficiency. The~No-U-Turn Sampler variant of HMC, now standard in frameworks like PyMC and Stan~\cite{Hoffman:2011ukg}, shows how algorithmic improvements can lead to widespread practical~adoption.

As cosmological analyses become increasingly sophisticated, understanding the advantages and limitations of these sampling methods becomes crucial. Many of these advanced methods show theoretical promise for next-generation cosmological surveys, where traditional sampling methods may become computationally prohibitive. However, their practical implementation often requires careful consideration of the specific problem context and available computational resources. This complexity underscores the importance of a detailed examination of various sampling approaches to develop more reliable numerical methods for cosmological~applications.

\textls[-15]{{{While our analysis demonstrates the varying strengths of different sampling methods, it also highlights that no single method is universally superior. The~choice of sampler should be guided by the specific requirements of the problem at hand, including parameter space dimensionality, likelihood characteristics, and computational~constraints}.}}

\newpage
\vspace{6pt}

\funding{This research was funded by Bulgarian National Science Fund grant number number~KP-06-N88/1..  
}

\dataavailability{The code will be made public upon publication on \url{https://github.com/dstaicova/samplers_benchmark} 
}


\conflictsofinterest{The authors declare no conflicts of interest.
}


%
%

\appendixtitles{yes} 
\appendixstart
\appendix
\section[\appendixname~\thesection. Accuracy~Estimates]{Accuracy~Estimates}\label{appendixA}

{\textbf Mean-based accuracy:}
For the mean-based accuracy, we compute:
\begin{equation}
    \text{Mean Error} = \sqrt{\frac{1}{d}\sum_{i=1}^d (\bar{x}_i - x_{\text{true},i})^2},
\end{equation}
where $d$ is the dimension, $\bar{x}_i$ is the mean of samples for parameter $i$, and~$x_{\text{true},i}$ is the true parameter value. This metric assesses how well samplers recover the true parameters, which is particularly relevant for point estimation tasks. For~weighted samples (as in nested sampling methods), we use the weighted mean $\bar{x}_i = \frac{\sum_{j=1}^N w_j x_{ij}}{\sum_{j=1}^N w_j}$, where $w_j$ are the sample~weights.

{\textbf Distribution-based accuracy:}
The distribution-based metric varies by test problem to capture the specific features of each distribution.

For the Gaussian case:
\begin{equation}
    \text{Gaussian Error} = -\frac{1}{2}\sum_{i=1}^d x_i^2
\end{equation}
which directly measures deviation from the zero-centered standard normal~distribution.

For the Rosenbrock function:
\begin{equation}
    \text{Rosenbrock Error} = \frac{1}{d-1}\sum_{i=1}^{d-1} \left[100(x_{i+1} - x_i^2)^2 + (1-x_i)^2\right]
\end{equation}
which measures how well samples follow the characteristic curved valley of the~distribution.

For the Gaussian mixture:
\begin{equation}
    \text{Mixture Error} = -\log\left(\exp\left(-\frac{1}{2}\sum_{i=1}^d (x_i+2)^2\right) + \exp\left(-\frac{1}{2}\sum_{i=1}^d (x_i-2)^2\right)\right)
\end{equation}
measuring the ability to sample from both modes at $x_i = \pm2$.

The final accuracy scores are normalized using fixed maximum error thresholds: 0.5 for Gaussian, 2.0 for Rosenbrock, and~5.0 for the mixture~model.

The mean-based metric is appropriate for unimodal distributions where point estimation is meaningful, while the distribution-based metric better captures performance on multimodal or highly curved distributions, where the mean alone may be~misleading.

\section[\appendixname~\thesection. Evidence~Computation]{Evidence~Computation}\label{sec:evidence}

The Bayesian evidence (marginal likelihood) $Z$ enables quantitative model comparison through Bayes factors $B_{12} = Z_1/Z_2$. Nested sampling algorithms compute $Z$ by transforming the multi-dimensional evidence integral into a one-dimensional integral over the prior volume $X$:
\begin{equation}
Z = \int L(\theta)\pi(\theta)d\theta = \int_0^1 L(X)dX,
\end{equation}
where $L(X)$ is the inverse of the prior cumulative distribution. The~algorithm maintains a set of ``live points'' drawn from $\pi(\theta)$ subject to a progressively increasing likelihood constraint $L(\theta) > L_i$. At~each iteration, the~lowest-likelihood point contributes $w_i = (X_{i-1} - X_i)L_i$ to the evidence sum, where $X_i$ is the remaining prior~volume.

For MCMC chains without direct evidence computation, approximate methods like MCEvidence~\citep{Heavens:2017afc} ({or the code based on it \url{https://github.com/yabebalFantaye/MCEvidence/blob/master/MCEvidence.py}}, {accessed on 16.01.2025}) estimate $Z$ through thermodynamic integration or other techniques. However, these approximations typically have larger uncertainties than nested sampling's direct~computation.

In our implementation, we use dynesty's built-in evidence calculation for nested sampling runs. For~comparison, MCEvidence can process the MCMC chains from traditional, HMC, and~slice sampling methods, though~with reduced precision compared to the nested~approach.

\section[\appendixname~\thesection. Implementation~Details]{Implementation~Details}
\label{sec:implementation}

For reproducibility, we detail the key implementation choices for each sampler:

\textbf{{Parameter Settings}}
\begin{itemize}
\item Traditional MCMC: 4 chains, 5000 draws (ndraws$\times$5) with 50\% burn-in, multiple adaptive step sizes (0.1, 0.05, 0.01) with tune\_interval=50
\item emcee: nwalkers = $\max(20 \times \text{ndim}, 40)$, nsteps=1000, 100 steps burn-in, $a=2.0$ for stretch move
\item HMC: 500 warmup steps, adapt\_step\_size=True, target\_accept\_prob=0.8, \linebreak  max\_tree\_depth=10
\item Slice: width parameter = 0.1, 4 chains
\item Nested (dynesty): 1000 live points, multi-ellipsoidal bounds, 'rwalk' sampling
\item PolyChord: nlive=100, num\_repeats = $\max(\text{ndim} \times 5, 30)$
\end{itemize}
\textbf{Priors:} For test problems, uniform priors on $(-5, 5)$ for all parameters in Gaussian, Rosenbrock, and~mixture~models.

The cosmological priors are:
 $\Omega_m \in (0.1, 0.5)$, $\Omega_K \in (-0.3, 0.3)$,
 $w \in (-2, 0)$, \linebreak  $w_a \in (-2, 2)$

\textbf{Likelihood Implementation}
The equations of the likelihoods are described in the main text. The~code uses \texttt{numpy} for standard computations and \texttt{jax.numpy} for gradient-based methods (HMC).

The marginalized BAO likelihood takes the form~\cite{Staicova:2021ntm, Benisty:2024lmj}:
\vspace{-6pt}\begin{equation}
\chi^2_{BAO} = C - \frac{B^2}{A} + \log\left(\frac{A}{2\pi}\right),
\end{equation}
where:
\vspace{-6pt}\begin{align}
A &= f^j(z_i)C_{ij}f^i(z_i), \\
B &= \frac{f^j(z_i)C_{ij}y^i_{obs}(z_i) + y^j_{obs}(z_i)C_{ij}f^i(z_i)}{2}, \\
C &= y^{obs}_j C_{ij}y^{obs}_i.
\end{align}

Here, $\vec{y}_{obs}$ is the vector of observed points at each $z$, and $f(z_i)$ is the model prediction. $C_{ij}$ is the covariance~matrix.

\textbf{Convergence Criteria}
All MCMC methods (traditional, emcee, HMC) use R-hat \linebreak  <$1.1$ and minimum effective sample size >$100$ per parameter. For~nested sampling, we use the default stopping criterion $dlogZ <0.1.$

\begin{adjustwidth}{-\extralength}{0cm}

\reftitle{References}




\isAPAandChicago{}%

%


\PublishersNote{}
\end{adjustwidth}
\end{document}